\documentclass[preprint,12pt]{elsarticle}

\usepackage{amssymb}

\journal{Journal of Quantitative Spectroscopy}

\begin{document}

\begin{frontmatter}



\title{
Simultaneous solution of Kompaneets equation and Radiative Transfer equation in the photon energy
range 1 - 125 KeV}

 \author[label1]{A. Peraiah}
 \author[label2]{M. Srinivasa Rao}
 \author[label2]{B. A. Varghese}
 \address[label1]{\# 57, 4th Cross, 36 main, BTM 1st stage, Madiwala, Dollar scheme, Bangalore 560068, India}
 \address[label2]{Indian Institute of Astrophysics, Bangalore 560034, India}



\begin{abstract}
Radiative transfer equation in plane parallel geometry  and Kompaneets equation is solved
simultaneously to obtain theoretical spectrum of 1-125 KeV photon energy range.
Diffuse radiation field is calculated using time-independent radiative transfer
equation in plane parallel
geometry, which  is developed using discrete space theory (DST) of radiative
transfer in a homogeneous medium  for different optical depths.
We assumed free-free emission and absorption and emission due to electron gas to be operating in the
medium. The three terms $n, n^2$ and $\displaystyle \bigg( {\frac {\partial n}{\partial x_k}}\bigg)$
where $n$ is photon
phase density  and $\displaystyle  x_k= \bigg( {\frac {h \nu} {k T_e}} \bigg) $, in Kompaneets
equation and those  due to
free-free emission are utilized to calculate the change in  the photon phase density in a
hot electron gas. Two types of incident radiation are considered: (1) isotropic radiation with the
modified black body radiation $I^{MB}$ [1]  and (2) anisotropic
radiation which is angle dependent. The emergent radiation at $\tau=0$ and reflected radiation
$\tau=\tau_{max}$ are calculated by using the diffuse radiation from the medium. The emergent and
reflected radiation contain the free-free emission and emission from the hot electron gas.
Kompaneets equation gives the changes in photon phase densities in different types of media.
Although the initial spectrum is angle dependent, the Kompaneets
equation gives a spectrum which is angle independent after several Compton scattering times.

\end{abstract}


\end{frontmatter}


\section{Introduction}

Compton and inverse Compton scattering play an important role in the processes of emission of X-ray
spectrum which has been observed in many compact astronomical objects.
Comptonization of X-rays  has been dealt mostly through Fokker-Planck approximation
(i.e, electron temperatures and photon frequencies are not high compared to the rest energy of the
electron)[2][3] [4] [5] and others. Primary non thermal X-ray and $\gamma$-ray emission
may reprocessed and end up in different energy band [6]. Several X-ray
observations obtained by Ginga do not confirm the power law spectral shape [7],
[8]. Spectral hardening have been noticed from Ginga satellite observations
[7] [9] [10] and  [11] of Seyfert galaxies and Cygx-1.

Analytical approximations of Compton reflection have been developed by
[12], on the basis of the separation of spacial
and energy transport in the transfer equation. This approach have the
advantage of facility to explore the model parameters. Later, many authors
[13] [14] [15] [16]  and [17] 
to implement Monte Carlo methods to treat several aspects of Compton
reflection models by taking more detail of the physics and geometry of the problem.
Burigana [18] studied the reflection problem with semi analytical transfer equation
and obtained the solution of the transfer equation.
Angle dependent Compton reflection of X-rays and $\gamma$-rays has been studied
by [19]. Studies on  Compton reflection of X-rays and $\gamma$-rays by [12] 
[20] [21] and others in shaping the emergent X-ray and $\gamma$-ray spectra.
Czerny \& Elvis [22], Zdziarski et al [23] and
Haardt \& Marasachi [24] studied the inverse Compton scattering in producing the high energy
photons in some of the compact sources in which low energy photons gain energy from hot electrons.
The geometry and distribution of matter in these objects would make the process anisotropic
[25] [6]  [26]. Haardt \& Marashi [24] and  Haardt  [27]
studied this process in plane parallel geometry.   Nishimura, Mitsadu \& Itoh [28] studied the same
problem but without angular dependence with isotropic scattering in the laboratory frame with
$kT_e<$ 150 KeV.

The escape probability and the energy distribution of the single scattered
photons are analytically calculated using Mellin integral transformations. Miyamoto  [29]
estimated the photon escape time distribution and effect of Comptonization on the spectrum.
Katz [30] obtained space distribution of the photons and spectrum changes by solving
Fokker-Planck equation from Shapiro et al [31]. The transfer of radiation estimated by using
Monte-Carlo method [32]. Another characteristic of X-ray spectra is its
variability over time. The time delay due to photon diffusion and the energy release and
multiple scattering of X-ray photons with thermal electrons due to comptonization would change the
nature of the emergent X-ray spectra. If $kT_e > h \nu$ then the photon gain energy and if
$kT_e < h \nu$ then the recoil effect will result in decrease in photon energy. Kompaneets
equation [33] allows us to find the time evolution of a given initial spectrum due to
comptonization in homogeneous medium. Illarionov \& Syunyaev  [34] obtained few analytical
solutions with a given initial spectrum in the soft X-ray region for continuum. Sunyaev \&
Titarchuk [25], Sunyaev \&Titarchuk [35]
calculated few solutions analytically for the time evolution of a given
initial soft X-ray spectra with Kompaneets equation and polarization of hard radiation. The
presence of electrons with random velocities would increase the energy of the photons by
$\displaystyle {\frac {\Delta \nu} {\nu} \approx {\frac {V^2} {c^2}} \approx
{\frac {kT_e} {mc^2}}}$. This
is the essential information that the Kompaneets equation Kompaneets [33]. Wehrse \& Hof [36]
have made line calculations using inverse Compton scattering.

The problem of the structure of a model atmosphere and the
spectrum formation under the influence of very hard X-ray external irradiation is not
satisfactorily solved in the present literature.
Recent papers with more general descriptions of Compton scattering were published by
many authors [37][38][39][40] [41][42][43] and [44] on
X-ray irradiated model stellar atmospheres.
Majczyna et al [45]had presented a model
which includes set of plane-parallel model atmosphere equations for a hot neutron star.
Suleimanov et al  [46][47]examined the effects of Compton scattering on the emergent spectra of DA white
dwarfs in soft X-ray range. They adopted two independent numerical approaches to the
inclusion of Compton scattering in the computation of pure hydrogen atmosphere in hydrostatic equilibrium.
The Kompaneets diffusion approximation formalism is used in one case,  cross-sections and
redistribution functions of [48]is used in another case. Chluba \& Sunyaev [49]
presented analytic solution of the Kompaneets equation for  low frequencies. They noticed that
multiple scattering of photons by hot electrons lead to brodening and shifting the spectral
features.
Exact stationary solutions
of the Kompneets kinetic equation is given by [50]. He also showed that the photon
input and output points always corresponds to finite frequencies.
Procopio \& Burigana [51] developed numerical code for the solution of Kopaneets equation and
discussed accuracy and applicability.

In this paper diffuse radiation field is calculated  to estimate the reflected and
the emergent radiation out of a plane parallel slab and also to  calculate the comptonization of the
photons with energy  between 1Kev-125Kev. Radiative transfer equation in plane parallel
approximation is solved with free-free absorption and emission and emission from hot electronic gas
in the medium. Photon phase density is estimated from the specific intensity derived from
the  equation of transfer for  both cases isotropic and anisotropic incident radiation.
Kompaneets equation is used to calculate time evolution of photon phase density.
At present we are not aiming to explain the observational results.

\section {\bf Description of the Theory:}

\noindent Kompaneets equation: The photon energy in an electron gas with random velocities,
would on the average increase by amounts approximately
$\displaystyle {\frac {\Delta \nu} {\nu} \approx {\frac {V^2} {C^2}} \approx
{\frac {kT_e} {mc^2}}}$ where k,
$T_e$, m, c, $\nu$ are the Boltzmann's constant, temperature of the electron gas, mass of the
electron, velocity of light, frequency of the photon respectively. The photon phase changes over
time due to scattering. This problem is well described by the Kompaneets equation which gives the
time evolution of a given initial photon spectrum due to Comptonization and  is given by
[33]  [1], and  [25],

\begin{eqnarray}
\bigg( {\frac {\partial n} {\partial y}\bigg)_c}={\frac {1}{x_k^2}} \frac {\partial}{\partial x_k}
\bigg[x_k^4\bigg(n+n^2+{\frac {\partial n}{\partial x_k}}\bigg)\bigg]
\end{eqnarray}
where
\begin{eqnarray}
x_k=\frac {h \nu} {kT_e}
\end{eqnarray}

\begin{eqnarray}
n=\displaystyle {\frac {c^2} {2h\nu^3} \bf I}
\end{eqnarray}

\begin{eqnarray}
y=\displaystyle \int^t_0  \frac {kT_e} {mc^2} \sigma_{T} {n_e} c dt
\end{eqnarray}

where $\bf I$ is the  specific intensity, h is the Planck constant, $\sigma_{T}$ is the
Thomson scattering
cross section and $n_e$ electron density. The $n^2$ term in equation (1) describes the induced
Compton interactions, ie., the energy being transfered from hotter photons to cooler
electrons as a result of which the photon energy is reduced  [52] and    [53].
The term is  important in the low frequency range. the term containing $n$
describes the drop in the photon energy $\displaystyle \frac {\Delta \nu} {\nu} \approx
{\frac {h  \nu} {mc^2}}$
in the photon energy through scattering which goes to the heating of the electrons. The term
containing
$\displaystyle \bigg( {\frac {\partial n} {\partial x_k}}\bigg)$ represents the diffusion of
photons mainly to increase the photon energy with consequent cooling of electrons.

Free-Free (Bremsstrahlung) processes influence the frequency redistribution of the photons
[33] and  [34] and is given by
\begin{eqnarray}
\displaystyle \bigg( {\frac {\partial n} {\partial y}}\bigg)_{ff}
=\frac {ke^{-x_k}} {\sigma x_k^3} \bigg \{1-n\bigg(e^{x_k}-1\bigg) \bigg\}
\end{eqnarray}
\noindent where the coefficient $a$ is given by
\begin{eqnarray}
a= \displaystyle {\frac {kT_e} {mc^2}} \sigma_T n_e c
\end{eqnarray}
and
\begin{eqnarray}
k&=&1.25 \times 10^{-12} g (x_k) n_e^2 T_e^ {-3.5} \nonumber \\
&=&k_0 g(x_k)
\end{eqnarray}
$g(x_k)$ being the Gaunt factor given by

\begin{equation}
g(x_k)= \left \{
\begin{array}{cc}
1  \quad \quad \quad\quad \quad \ \ for \ \ x_k > 1 \\
       \displaystyle {\frac {\sqrt {3}} {\pi}} ln \bigg( \frac {2.35} {x_k}\bigg) \ \ for \ \ x_k<1
\end{array}\right.
\end{equation}
Therefore the complete equation which includes both free-free and Compton processes is given by
\begin{eqnarray}
\frac {\partial n}{\partial y}&=&\bigg (\frac {\partial n}{\partial y}\bigg)_c+
\bigg (\frac {\partial n}{\partial y}\bigg)_{ff} \nonumber \\
&&=\frac {1} {x_k^2} {\frac {\partial}{\partial x_k}} \bigg \{x_k^4\bigg(n+n^2+ \frac {\partial n}
{\partial x_k}\big) \bigg\} \nonumber \\
&&+ \frac {ke^{-x_k}} {ax^3_k} \bigg \{ 1-n\bigg(e^{x_k}-1\bigg)\bigg\}
\end{eqnarray}

\noindent Equation (9) is solved numerically on discrete mesh of frequency points. We obtain the
spectrum of occupation numbers of photons in the phase space as $n$ recursive relation given by

\begin{eqnarray}
n_{i+2}&=&\bigg(p_tA_6\bigg)^{-1} \bigg[ n_i\bigg(\alpha_1-1\bigg)+ n_{i+1}\bigg(1-\alpha_2\bigg)\nonumber \\
&&-p_t\bigg\{A_{10}+M\bigg(n_i, n_{i+1}\bigg)\bigg\}\bigg]
\end{eqnarray}
\noindent where
\begin{eqnarray}
\alpha_1=p_t\bigg(-2A_1+A_4+4A_5-A_6+A_{11}\bigg)
\end{eqnarray}
\begin{eqnarray}
\alpha_2=p_t\bigg(2A_1+A_4+4A_5-2A_6-A_{11}\bigg)
\end{eqnarray}
\begin{eqnarray}
{\bf M} \bigg(n_i, n_{i+1}\bigg)=n_i^2\bigg(A_1-A_4\bigg)+n^2_{i+1}\bigg(A_1+A_4\bigg)
+2A_1n_in_{i+1}
\end{eqnarray}
where
$$
A_1=\frac{1}{2} \bigg(x_k^{i+1}+x_k^i\bigg);
\quad A_2=A_1^2;
\quad A_3=\bigg(\Delta x_k\bigg)^{-1};
$$
$$
A_4=A_2 A_3;
\quad \quad \quad A_5=A_1 A_3;
\quad A_6=A_2 A_3^2;
$$
$$
A_7=A_1^3;
\quad \quad \quad A_8=e^{-A_1};
\quad \quad \quad A_9=e^{A_1}=\bigg(A_8\bigg)^{-1}
$$
$$
A_{10}= FA_5A_1^{-3};
\quad \quad A_{11}=\frac {1}{2} A_{10}\bigg(A_9-1\bigg)
$$
$$
p_t=\Delta t .{\bf P};
\quad {\bf P}= \frac {kT_e \sigma_T n_e}{mc}=3.37 \times 10^{-24} T_e n_e
$$
$$
{\bf F}= 3.7147\times 10^{11} g(x) n_e T_e^{-4.5}
$$
\begin{eqnarray}
x_k=\frac {h \nu}{kT_e}=\frac {5.92648\times10^9}{T_e} x; \quad x=\frac {h \nu}{mc^2}
\end{eqnarray}
\subsection {\bf Solution of  Radiative Transfer Equation}

We obtain the initial spectrum of n in the equation (3) by solving the equation of transfer for
${\bf I}$, the specific intensity is given by [54], [55] and
[2] and others
\begin{eqnarray}
\frac {1}{c} \frac {\partial I_{\nu}}{\partial t}+{\bf I} \frac {\partial I_{\nu}}{\partial r}
=j_{\nu}-n\sigma_{\nu} I_{\nu}+ \displaystyle \frac {1}{c}  \bigg( {\frac {\partial I_{\nu}}
{\partial t}}\bigg)_{scat}
\end{eqnarray}
where $I_\nu (r, l, t) $ is the specific intensity of radiation in \linebreak
$ergs \  cm^{-2} S^{-1} {H_z}^{-1} {str}^{-1}$,
at position $r$, time $t$, and frequency $\nu$ traveling in the direction given by the unit vector
$\bf I$. In the above equation, it is convenient to use the photon energy $\epsilon=h\nu$ rather
than frequency. $(I_\nu)$ can be written as $I( e)= \displaystyle {\frac {I_{\nu}}{h}} $
(KeV $cm^{-2}S^{-1} KeV^{-1} {str}^{-1}$). The quantity $n\sigma_{\nu}$=$\displaystyle
\sum_j n_j \sigma_j(\nu)$
is the absorption coefficient, $n_j$ is the number of absorbers of type j and $\sigma_j(\nu)$ is the
total cross section of true absorption of photon energy. All effect of electron scattering are included
in the term $\displaystyle \bigg( \frac {\partial_I} {\partial t}\bigg) _{scat}$.
These were studied by  [54]
which are valid for $1 KeV \le\epsilon\le1GeV$ for which the photon occupation number n is less than unity
which is valid for many X-ray and $\gamma$-ray sources. If $n <<1$  then the induced scattering
is neglected.
In a cold electron gas for $\epsilon \ge 1 $KeV, [54] gives,
\begin{eqnarray}
\bigg( \frac {\partial_n} {\partial t}\bigg) _{scat}= \displaystyle cn_e \int d\omega \bigg(
\frac {d\sigma} {d \omega}\bigg)
\bigg[n\bigg(r,\Gamma,t\bigg)-n\bigg(r, \Gamma^\prime,t\bigg)\bigg]
\end{eqnarray}
correct to the lowest $\displaystyle \frac {pc}{\epsilon}$ and $\displaystyle \frac {p}{mc}$
where $p$
is the momentum of the targeted
charged particle in the observers frame of reference, $\displaystyle \Gamma=\frac {\epsilon}{c}$
is the photon
momentum
before scattering and $\Gamma^{\prime}$ is the momentum after scattering  [55].
These are related as
\begin{eqnarray}
\frac  {\Gamma^{\prime}}{\Gamma}= \frac {1} {{1+\bigg( \frac {\Gamma}{mc}\bigg)}\bigg(1-\cos \theta\bigg)}
\end{eqnarray}
where $\theta$ is the angle between the vector moments of $\Gamma$ and $\Gamma^{\prime}$ and
$\displaystyle \frac {d\sigma}{d \Omega}$ is the differential cross section for Compton
scattering in the observer's frame
where the plasma is at rest. This is given by Klen-Nishana formula from [56] and [57]
\begin{eqnarray}
\frac {d\sigma}{d \omega}= \frac {1}{2} r_0^2 \bigg( \frac {\Gamma^{\prime}}{\Gamma}\bigg)^2
\bigg(\frac {\Gamma^\prime} {\Gamma}+{\frac {\Gamma}{\Gamma^\prime}}-sin^2\theta\bigg)
\end{eqnarray}
where $r_0=\displaystyle {\frac {c^2}{mc^2}}=2.85 \times 10^{-13}$ cm is the classical radius of the electron.
For an incoming photon energies $\epsilon^\prime$ small compared to the $mc^2$, which is seen
from the equation (17) that $\displaystyle \frac {\Delta \epsilon^\prime}{\epsilon} < < 1$. The collision
term $ \displaystyle \bigg( \frac {\partial n} {\partial t} \bigg)_{scat}$ is then given by
Fokker-Planck type expression
derived from (16) [58] and [59]. The resulting equation of
transfer is given by (to the lowest order of $\displaystyle {x=\frac {\epsilon}{mc^2}
=\frac { h\nu}{mc^2}}$) [55].
\begin{eqnarray}
\frac {1}{c} \frac {\partial \delta I_x} {\partial t} + {\bf I} \frac {\partial I_x}{\partial r}+
n_e \sigma_T \sigma(x) I_x=j_x + \sigma_T n_e x \frac {\partial} {\partial x}
{\bigg(x {\bf I}_x \bigg) }
\end{eqnarray}
where $n_e\sigma_T \sigma_(x)$ is mass absorption coefficient, $j_x$ is the emissivity

in $ergs\  cm^{-3} s^{-1} Hz^{-1} str^{-1}$.

\noindent And
\begin{eqnarray}
\sigma (x)=\sigma_s(x)+\sigma_a(x)+\sigma_a^{ff}(x_k)
\end{eqnarray}
where for $\displaystyle h\nu <<{\frac {1}{4}} mc^2$ [56], the quantities
$\sigma_x(x), \sigma_a(x)$ are given by
\begin{eqnarray}
\sigma_s(x_k)= 1-3x+\frac {94}{10} x^2-28 x^3+ \frac {552}{7} x^4-\ldots
\end{eqnarray}
\begin{eqnarray}
\sigma_a(x)= x-\frac {42}{10} x^2 + \frac {147}{10} x^3-\frac {1616}{35} x^4+\ldots
\end{eqnarray}
In the present analysis we restrict our study to the energy range of 1 KeV-125 KeV
$\displaystyle ({h \nu}< \frac {1}{4} {mc^2})$.
The free-free absorption coefficient $\sigma^{ff}_{a}{(x)}$
is given by  [1]
\begin{eqnarray}
\sigma^{ff}_{a}(x_k)=1.79 \times10^{-18}T_e^{-\frac {3}{2}} Z^2 n_i x^{-2} {\bar g}_{ff}
\end{eqnarray}
where $z$ is the charge and $n_i$ is number density of ions and ${\bar g}_{ff}$ is the
Gaunt factor which is of the order of unity. The quantity $j(x)$ is the emissivity is given by
\begin{eqnarray}
j(x)&=&\frac {1}{4} {\bf B}^{ff}_{em}(x_k)+B_e(x,\beta) \nonumber \\
&&+ \frac {1}{4\pi} \int
R(x,x^{\prime})I(x^\prime, \mu^\prime) dx^\prime d\mu^\prime
\end{eqnarray}
where ${\bf B}^{ff}_{em}(x)$ is the Bremsstrahlung emission given by [1]
\begin{eqnarray}
{\bf B}^{ff}_{em}(\nu)=6.8 \times 10^{-38} \ Z^2 n_i n_{e} \ T^{-\frac {1}{2}}_{e} \
e^{-\frac {h\nu}{kT}} \ {\bar g}_{ff}
\end{eqnarray}
\begin{eqnarray}
{\bf B}^{ff}_{em}(\nu)&=&6.8 \times 10^{-38} \ Z^2 n_i n_{e} \times \nonumber \\
&& T^{-\frac {1}{2}}_{e}
\ exp  \bigg [ -{\frac{5.926\times10^9}{T_e}} x\bigg] \ {\bar g}_{ff}
\end{eqnarray}
Further the quantity ${\bf B} (x,\beta)$ is given by [1], [27],
and [28]

\begin{eqnarray}
&&{\bf B}_e(x, \beta)\nonumber\\
&&= {\displaystyle \frac {n_e \sigma_T {\bf I}\delta(x-x_0)}{4\gamma^2 \beta^2 \epsilon}}
\nonumber\\
&& \times
\left \{
\begin{array}{ccc}
(1+\beta) {\displaystyle \frac {x_0}{x}} -(1-\beta)\quad   for \ \ \   {\displaystyle \frac {1-\beta}{1+\beta}<\frac {x_0}{x}<1}  \\
(1+\beta) -{\displaystyle \frac {x_0}{x}}(1-\beta) \quad for  \ \ \ {\displaystyle 1< \frac {x_0}{x}< \frac {1+\beta}{1-\beta}}  \\
0   \quad \quad \quad\quad otherwise \\
\end{array}\right.\nonumber\\
&&
\end{eqnarray}
where ${\bf I} \delta(x-x_0)$ is the monochromatic specific intensity, $\displaystyle
\beta=\frac {v}{c}$
and $\displaystyle \gamma =(1-\beta^2)^{\frac {-1}{2}}$. ${\bf B}_e(x,\beta)$ becomes
effective only at higher
temperatures as the energy is transformed from hot electron gas to the cold photon gas. We need
to study the redistribution of energies at different frequencies. This is given by [55],
and [56]

\begin{eqnarray}
&&R(x, x^\prime)= \nonumber \\
&&\frac {3}{8} \frac {\sigma_Tn_e} {x^2}
\bigg[\frac {x^\prime}{x}+\frac {x}{x^\prime}-2 \bigg(\frac {1}{x^\prime}-\frac {1}{x}\bigg)
+\bigg(\frac {1}{x^\prime}-\frac {1}{x}\bigg)^2\bigg]
\end {eqnarray}
where $x, x^\prime$ are the energies of photons before and after the scatterings.
Equation (19) is valid for $\displaystyle h\nu<<mc^2$ [55].
The last term on the right hand side of
equation (19) is the same as the term containing $n$ in the Kompaneets equation (1). Therefore
equation (19) can not fully represent the diffusion of photon energy which is adequately
done by Kompaneets equation as this contains the $\displaystyle \frac {\partial n}{\partial x}$
and $n^2$ term
also. We need to solve equation (1) and (19) simultaneously without the last term on the RHS of
equation (19). Therefore the time independent plane parallel equation of radiative transfer
is written
as
\begin{eqnarray}
&&\mu \frac {dI(x, \mu, z)}{dz}+n_e\sigma_T \sigma(x, T) I(x, \mu, z) \nonumber \\
&=&\sigma_a(x) {\bf B}_e(x,\beta) +\frac {\sigma^{ff}(x, T_e)}{4\pi} {\bf B}^{ff}_{e} (x,T_e)\nonumber \\
&+& \frac {\sigma_s} {8\pi} \int^{\infty}_{0} \int^{+1}_{-1} R(x, x^\prime) I(x^\prime,\mu^\prime)
d\mu^\prime
\end{eqnarray}
where $\mu$ is the cosine of the angle made by the ray with normal to the plane parallel layers,
important steps of the method is given Appendix  and

\begin{eqnarray}
\sigma^{ff}=n_e \sigma_T \sigma_{a}{^{ff}}(x, T_e)=n_{\epsilon} \sigma_T\sigma_a^{ff}.
\end{eqnarray}

\section {Result and Discussion}

\subsection{Boundary conditions}

We shall give the boundary condition of the incident radiation as follows:

\noindent Case 1:
\begin{eqnarray}
{\bf I}(\tau=T,\mu)= {\bf I}^{MB}
\end {eqnarray}
and
\begin{eqnarray}
{\bf I}(\tau=0, \mu)=0
\end {eqnarray}
where
\begin{eqnarray}
{\bf I}_{\nu}^{MB}=  8.4 \times 10^{-4} T^{\frac {5}{4}}_{e} \rho^{\frac {1}{2}}
{\bar g}^{\frac {1}{2}}_{ff} x_k^{\frac {3}{2}} c^{\frac {-x_k}{2}}\bigg( c^{x_k}-1\bigg)
^{\frac {-1}{2}}
\end {eqnarray}
is the modified black body radiation given in [1]. Here
$T_{e}$ is the temperature, $\rho$ is the density, ${\bar g}_{ff}$ is the
Gaunt factor given by
\begin{equation}
{\bar g}_(ff)= \left \{
\begin{array}{cc}
1  \quad \quad \quad\quad \ \ \ for \ \ x_k > 1 \\
       \displaystyle {\frac {\sqrt {3}} {\pi}} ln \bigg( \frac {2.35} {x_k}\bigg) \ \ for \ \ x_k<1
\end{array}\right.
\end{equation}
\begin{eqnarray}
x_k=\frac {h \nu}{kT_e}=\frac {5.92647\times 10^9}{T_e} x; \quad x=\frac {h \nu}{mc^2}
\end{eqnarray}

\noindent Case 2:

\begin{eqnarray}
\quad {\bf I}(\tau=T, \mu)={\bf I}^{MB} \times \frac {6}{(3+2b)}\ (\mu+b\mu^2)
\end{eqnarray}
with $b$=2 
and
\begin{eqnarray}
{\bf I}(\tau=0, \mu)=0
\end {eqnarray}

in this case we set ${\bf B}_e(x, \beta)$=0 but free-free emission is included in the medium.

The solution of transfer equation is developed  using DST and obtained the solution. The stability and
the accuracy of the method is  checked through the following;
We divide the medium into a
number of "Cells"  whose thickness is less than or equal to the critical $(\tau_{crit})$.
The critical thickness is determined on the basis of physical characteristics of the medium.
$\tau_{crit}$ ensures the stability and uniqueness of the solution and other steps
are mentioned in Appendix.
By considering above,  twenty trapezoidal points for energy,
four angle points and fifty plane parallel
homogeneous layers are considered in our calculations.
As there are many parameters to be studied we choose few representative of these
to highlight the physical processes. We have chosen temperature $T_e=5 \times 10^9$ K
and $10^{11}$ K, $n_e=10^{15} {cm}^{-3}, \tau=1$  with the initial conditions given in equation
(31) to (37). We assumed a purely ionized hydrogen gas so that $\rho$ in equation (33) becomes
$n_e m_p$ where $m_p$ is the mass of the proton.
The energy points$\displaystyle (x=\frac {h\nu}{mc^2})$ are chosen from 0.0002 to 0.24 on
trapezoidal quadrature
corresponding approximately to 1 KeV and 125 KeV respectively. The four angle points are chosen
on the Gauss-Legendre quadrature on (0, 1) with $\mu_1$=0.6943, $\mu_2$=0.33001, $\mu_3$=0.66999,
$\mu_4$=0.93057 and the corresponding weights are $C_1$=0.1739, $C_2$=0.3261, $C_3=C_2$, $C_4=C_1$.

\noindent Case I: The incident radiation field represented by  $I^{MB}$ the modified black body
radiation (ergs cm$^{-2}$ S$^{-1}$ Str$^{-1}$), is plotted in figure 1 for energy points at temperature
5$\times$ 10$^9$, 10$^{10}$, 10$^{11}$ K 
to see how a photon of given energy would get transformed
through the medium for different temperatures. 

In figure 2, We plotted the variation of $I^-$, $I^+$ and
their corresponding
photon phase densities $n^+$ and $n^-$ with energy for the parameters are shown in the figures
(see equation A. 20, A. 21, and 3) for $\mu_4$  
when   initial intensity is given at
$I^-(N=50)=1$ and free-free emission is not included.
Figure 2(a) shows the emergent intensity and figure  2(b)
gives reflected intensity. Figure 2(c) and 2(d) give their corresponding photon phase densities
N=50, 25, 1 corresponding to $\tau=$0.5 and $\tau=$0. 
We notice that,  a  small fraction of emergent intensity  is reflected at $N=50$
and emergent intensity $I^-$ increases marginally at higher energy points. 

In figures 3(a, b c, d) we show the variation of $I^-(\mu)$, $I^+(\mu)$,
$n^-(\mu)$,
$n^+(\mu)$ across the medium for the photon energy 100 KeV.
$I^-(\mu$ at $N=50)$ increases as it acquires more
free-free emission in the medium. $I^-(\mu_1)$ shows maximum values at about N=40 and falls slowly towards
N=1.
$I^+(\mu_1) > I^+(\mu_2) >I^+(\mu_3) >I^+(\mu_4) $ at all layers. The intensities fall sharply from N=10
reaching minimum at $n=1$.

Figures 4(a, b, c, d) contain the information of $I^-$, $I^+$, $n^-$, $n^+$ with respect to $\epsilon$ at
$T_e$=10$^{10}$ K. The incident intensity $I^-$ at N=50 is given as $I^{MB}$ and free-free emission
from the medium is assumed. 
$I^-(\epsilon)$ reduces toward higher energy points although there is a slight increase
at $\epsilon \approx $1-10 KeV. $I^-$ at $N=1$ that is at the emergent point is considerably reduced
compared to the that at $N=50$. It appears that free-free emission at $T_e$ =10$^{10}$ K is not
effective
in contributing to $I^-$ at higher energies. The reflected intensities $I^+$ at $N=50$ is higher than that
at $N=1$ which shows that more energy  is reflected at $N=50$ than that at $N=1$ at the emergent side $(\tau=0)$.
At $N=50$ , $I^-$ is greater than $I^+$ by a factor of 10$^4$ while at $N=1$, $I^-$ is larger by a
factor of
10 at lower energies and $I^- \approx I^+$ at higher energies. The $n^-$ and $n^+$ given in 4(c) and 4(d)
reflect the same variations shown figures 4(a) and 4(b) respectively. 

Figures 5(a) and 5(b) contain the
transfer of an 8 KeV photon energy through the medium from $N=50$ to $1$. $I^- (8 KeV, \mu_1)$ falls rapidly
between
$N=50$ and $N=1$, that is from $\tau=\tau_{max}=1$ and $\tau=0$, changing by a factor of 10$^4$ while
$I^-(8 KeV, \mu_2)$ change by a factor of 10 and $I^-(8 KeV, \mu_3, \mu_4)$ change only by a factor of
approximately 2. The situation with $I^+$ is different in that $I^+(8 KeV, \mu_1)$ change is not as
steep as those of $I^+(8$ KeV, $\mu_2, \mu_3, \mu_4)$. All $I^+$ (8 KeV) fall rapidly in the
extremely outer
layers from $N\approx 10$ to $N=1$. A similar variation can be seen in figures 5 (c) and 5(d) for
$n^-$ and $n^+$ corresponding to $I^-$ and $I^+$.  

Figure 6(a, b, c, d)  shows the enhancement of photon phase space density given by the recursive
relation given in equation (10) for temperature $T_e=10^{10} K, \tau=1$ and the emergent intensities
$I^-(N=1)$, with $I^-(N=50)= I^{MB}$ and free-free emission.
We have calculated the enhancement
of ${n}^-$ for different values of $\Delta t$(see equation 13). In the recursive relation of (10),
we choose the first two energy points ${n}_1^-$ and  ${n}_2$ from $I^-(N=1, \epsilon=\epsilon_1)$
$I^-(N=1, \epsilon=\epsilon_2)$ respectively. The photon phase space densities
in figure 6, (for $\mu_1$ and $\mu_4)$) rise sharply from $\Delta t=100$. It is interesting to note
that ${n}^-(\mu_1)$ are much smaller than those of ${n}^-(\mu_4)$ up to $\epsilon \approx$ 50 KeV.
Above 50 keV, ${n}^-(\mu_4)$ and ${n}^-(\mu_1)$ coincide and the phase space densities become angle
dependent.

\noindent Case II: Calculations are done with  anisotropic boundary conditions given in
the equations (36) \& (37) due to [25]. Figure 7(a, b) represents $I^-$ and $I^+$
for $\tau=1$ for the boundary condition

$$I^-(\tau=\tau_{max})=I^{MB} \times \frac {6}{(3+2b)} (\mu+b \mu^2)$$
\noindent and
$$I^-(\tau=0, \mu)=0 $$

with $b$=2 and $T=10^{10}$ K. We assumed free-free emission in the medium. No emission from the hot
electron gas is given in the medium. There appears to be a maximum at about 8-9 KeV and a steep fall
in the energy range of $\epsilon > $ 10 KeV, More  energy seems to be emerging through $\mu_1$ than
through $\mu_4$, particularly at N=1. The reflected $I^+$ is
considerably reduced and is flat irrespective of energy range. 
Figure 8(a, b)  give  the result of $I$'s
for $\tau=1$ with the initial condition

$$I^-_\nu(\tau=\tau_{max}, \mu)=I^{MB} \times 2 \mu \ \ ln \bigg( \frac {1+\mu}{\mu}\bigg) $$

\begin{figure}
\centering
\includegraphics [width=8cm,height=8cm] {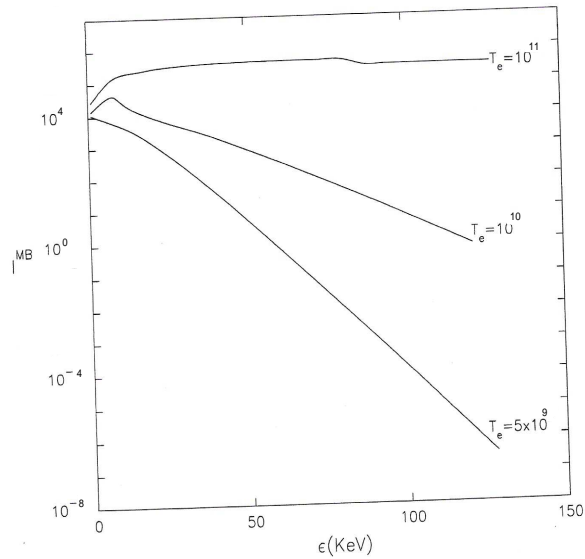}
\caption {$I^{MB}(in \ \ ergs \ cm^{-2} S^{-1} HZ^{-1} Str^{-1})$ versus photon energy $\epsilon$ in KeV,
for temperatures $T_e$= 5 $\times$ 10$^9$K, 10$^{10}$ K, 10$^{11}$K }
\end{figure}

\begin{figure*}
\centering
\includegraphics [width=15cm,height=17cm]  {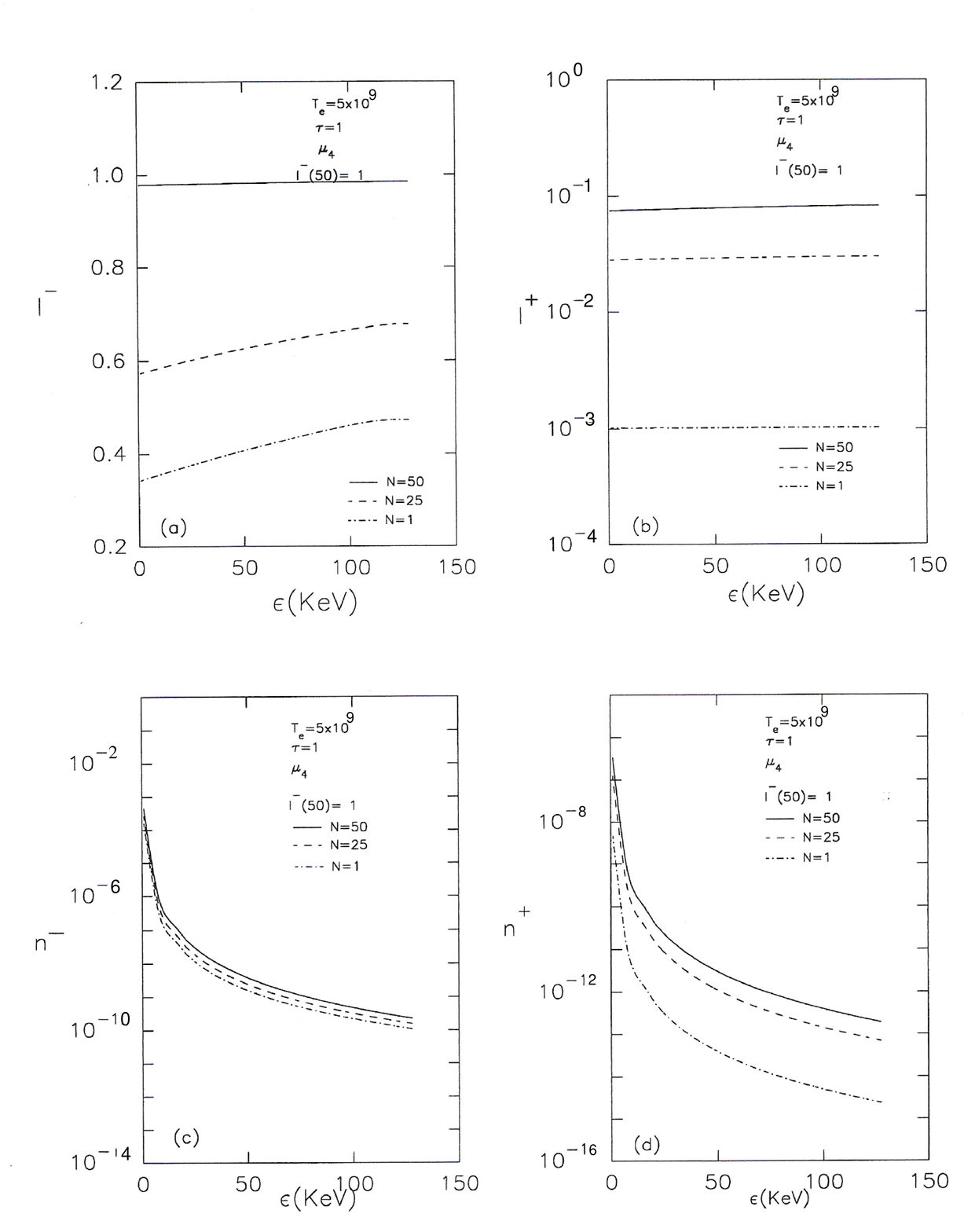}
\caption { (a) $I^-$ the emergent intensity (in the direction of $\tau \rightarrow 0$) at different layers at
$N$=50 ($\tau=\tau_{max}$), $N$=25($\tau=0.5$) and $N$=1 ($\tau=0$) for angle $\mu_1$ for temperature
$5 \times 10^9$ K. (b) $I^+$ the reflected intensities at N=50, 25, and 1. 
The initial incident intensity is $I^-(N=50)=1$. (c) and (d):
shows the photon phase densities $n^-$ and $n^+$ corresponding to $I^-$ and $I^+$ given in figures (a)
and (b) respectively}
\end{figure*}

\begin{figure*}
\centering
\includegraphics [width=14cm] {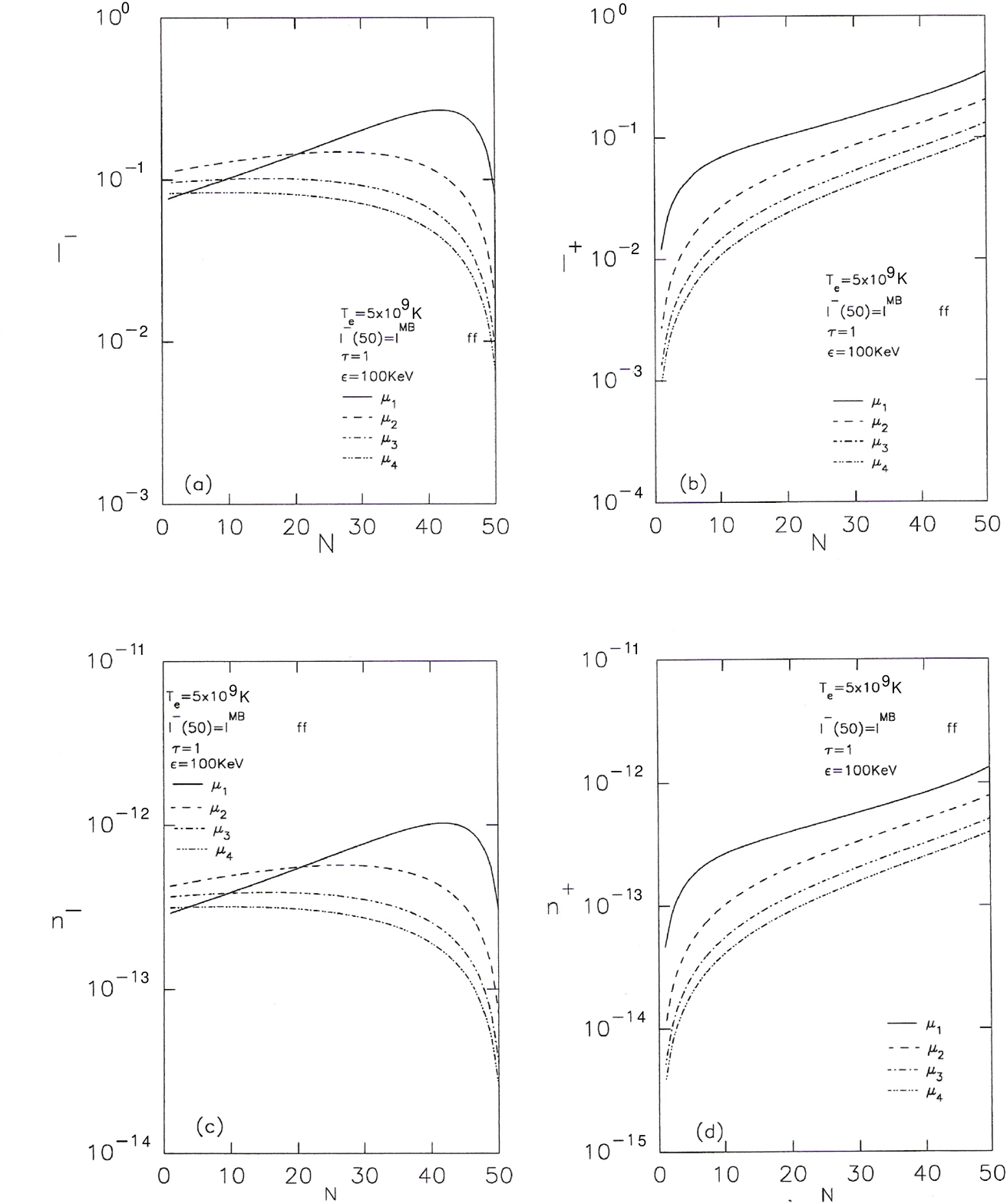}
\caption { The transfer of photon with $\epsilon$=100 KeV energy across the medium from $N=50$ to $N=1$,
in a medium with free-free emission and $I^-(N=50)=I^{MB}$ as the initial intensity.}
\end{figure*}

\begin{figure*}
\centering
\includegraphics [width=15cm] {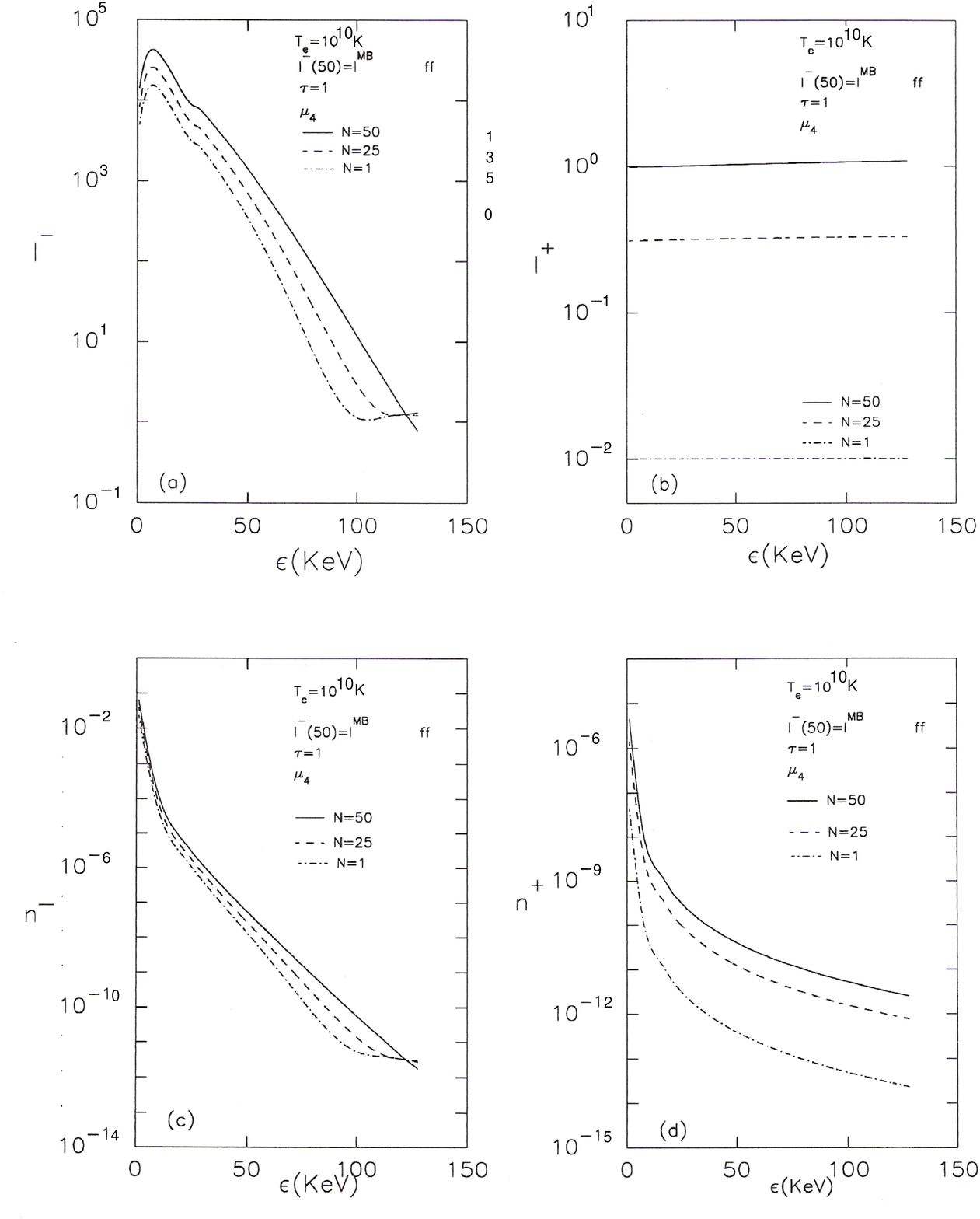}
\caption { Same as those given in figure 3, but with $T_e$=10$^{10} K$ }
\end{figure*}

\begin{figure*}
\centering
\includegraphics [width=14cm] {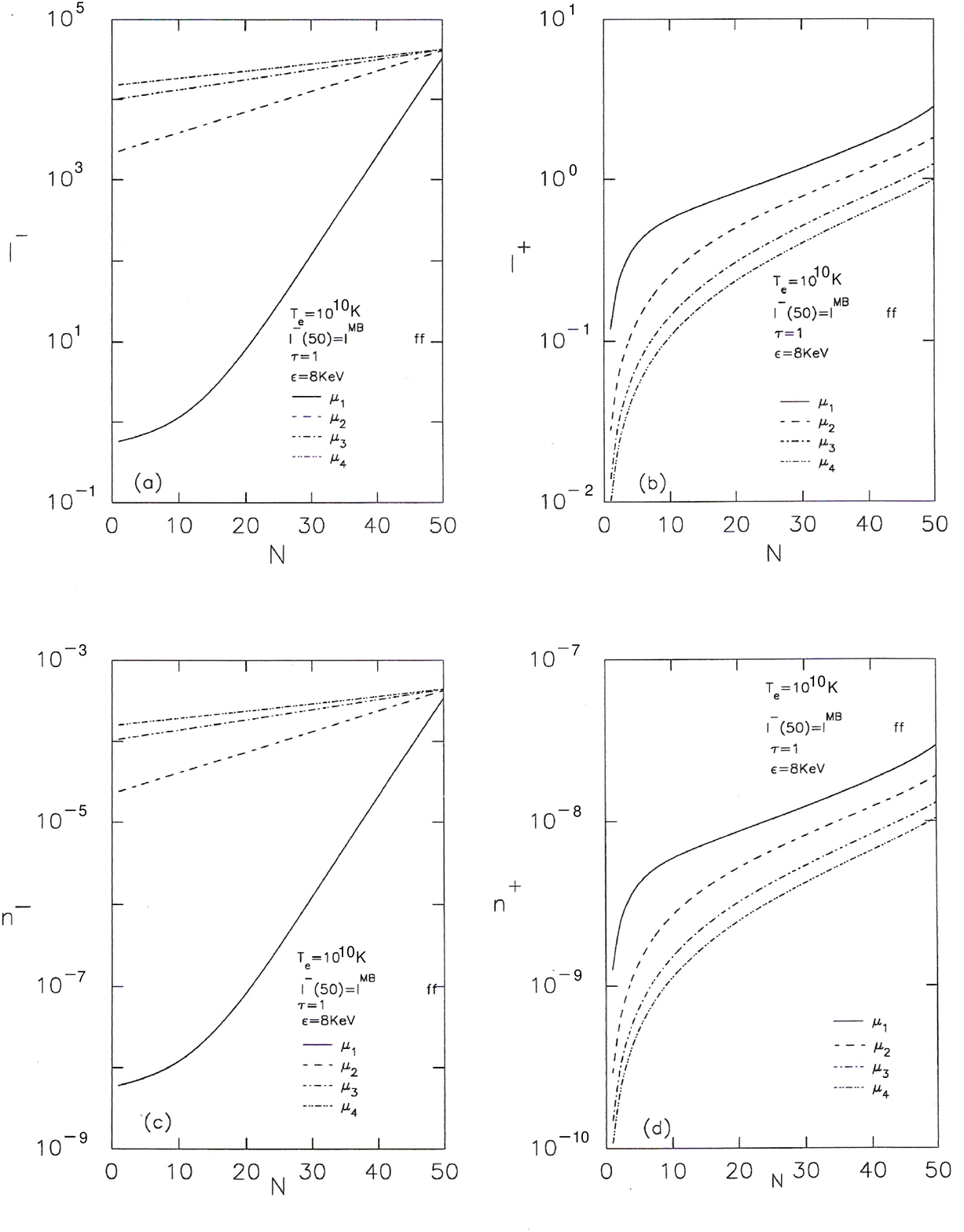}
\caption { The transfer of photon with $\epsilon$=8 KeV energy across the medium from $N=50$ to $N=1$,
in a medium with free-free emission and $I^-(N=50)=I^{MB}$ as the initial intensity for  $T_e$=10$^{10} K$ }
\end{figure*}

\begin{figure*}
\centering
\includegraphics [width=15cm,angle=0] {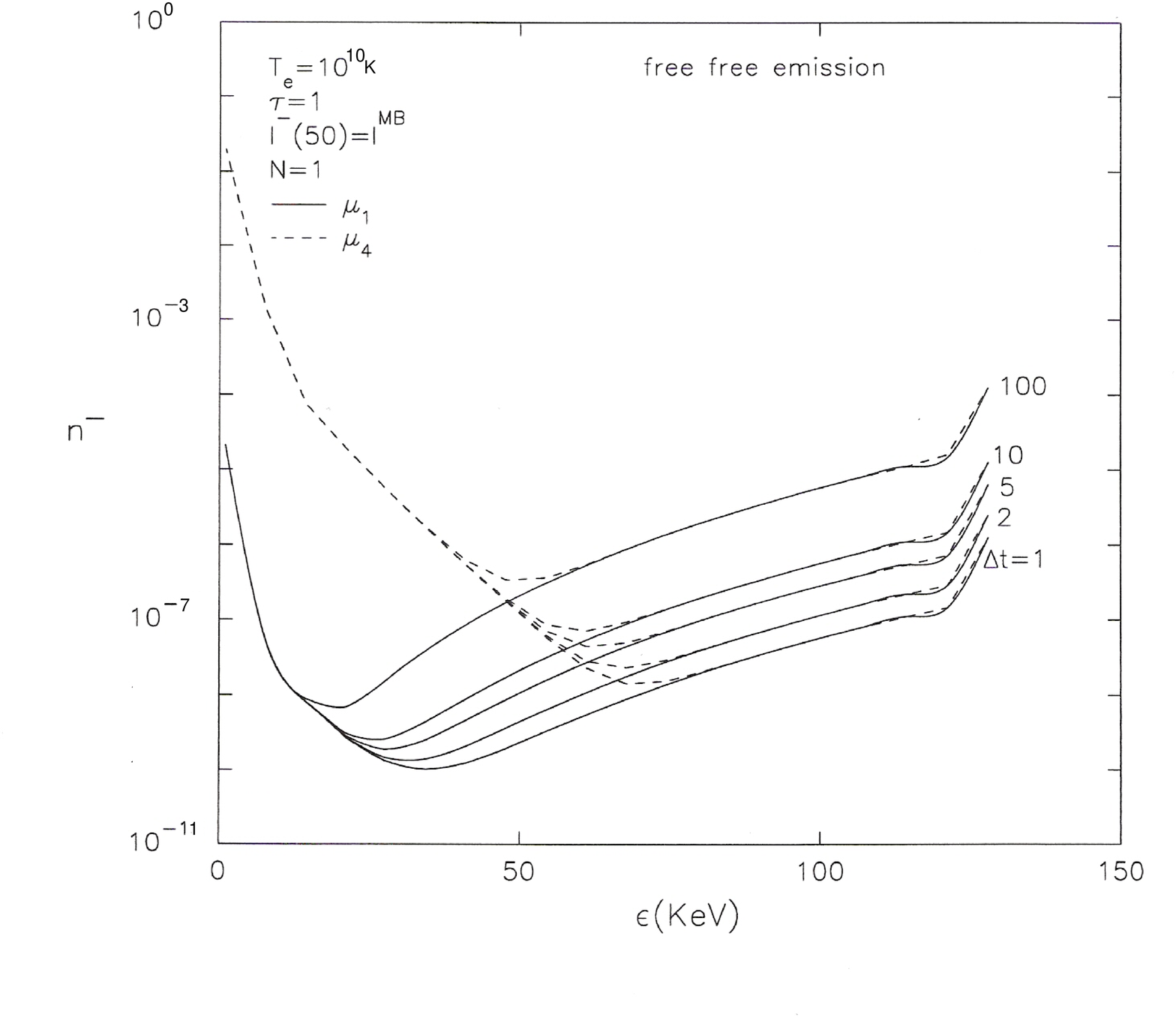}
\caption { The photon phase density $n^-$ corresponding to $I^-$ is given versus $\epsilon$. The enhancement
of $n^-$ for $\delta t$=1, 2, 5, 10, 100 is shown. The incident radiation is $I^-(N=50)=I^{MB}$ in a medium
with free-free emission  }
\end{figure*}

\begin{figure*}
\centering
\includegraphics [width=15cm,angle=0] {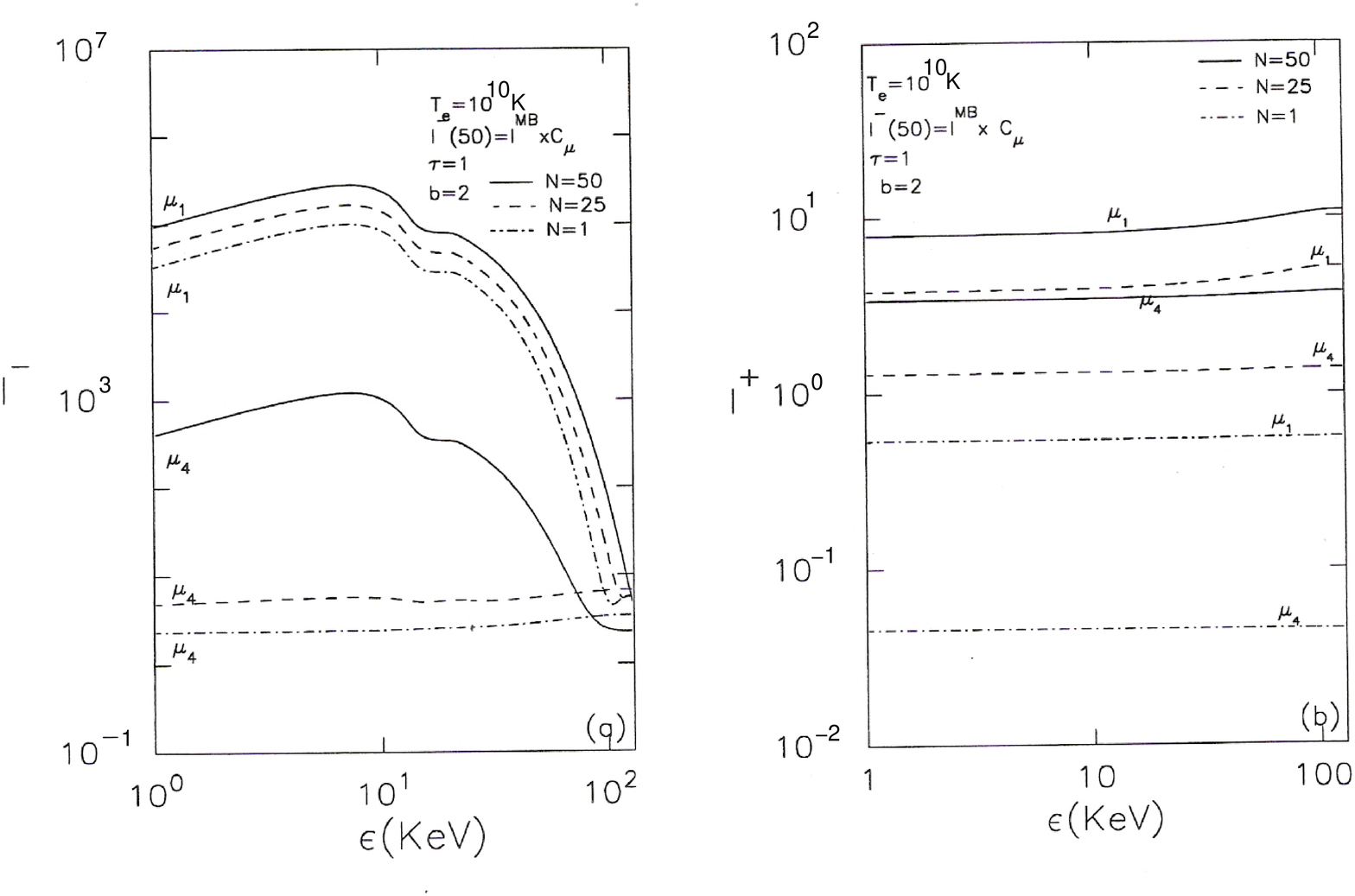}
\caption { The emergent and reflected intensities $I^-$ and $I^+$ with initial condition $I^-(N=50)=
I^{MB} \times C_\mu $where $C_\mu= \displaystyle {\frac {6} {3+2b}} \bigg(\mu+b \mu^2\bigg)$}
\end{figure*}

\begin{figure*}
\centering
\includegraphics [width=15cm,angle=0] {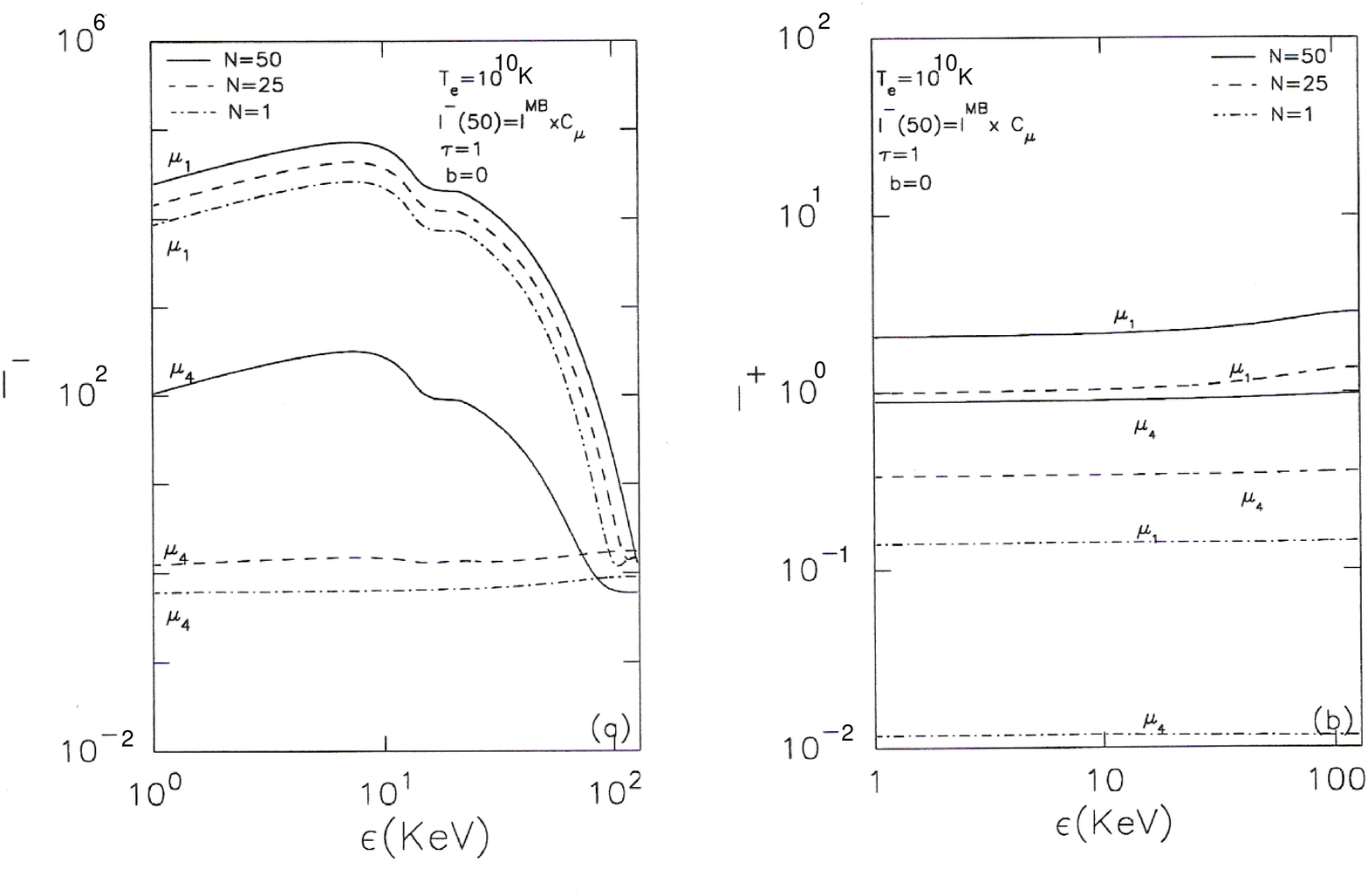}
\caption { Same as those given in figure 7  with $I^-(N=50)=I^{MB} \times D_\mu$ where $D_\mu=
2 \mu \ ln \ \bigg( \frac {1+\mu}{\mu}\bigg)$ }
\end{figure*}

\section {Conclusions}

Using discrete space theory of radiative transfer comptonization spectra is calculated in a plane
parallel atmosphere
with isotropic and anisotropic as a incident radiation.  The accuracy of the solution is assumed by
observing 1)strict radiant flux conservation 2) non-negativety and continuity of the solution across
$(-1 \le \mu \le 1)$ by taking the step size for the democratization.
We estimated the photon phase densities from the
specific intensities which is calculated from radiative transfer equation and the time evolution
of photon phase density by employing Kompaneets equation.
There is substantial differences are noticed in the nature of emergent intensity variations with  free-free
emissions and absorption cases. We also studied when
a photon of given energy would get transformed through the
medium and noticed that when transfer takes place radiation field becomes angle dependent.
We noticed that the initial spectrum is angle dependent and the Kompaneets equation gives a angle
independent spectrum after many Compton scatterings.
We are plan to extend the  above method for spherically symmetric
approximation to study the stellar atmospheres in X-ray binaries and accretion discs atmospheres
in active galatic nuclei(AGN) etc.

\section {Acknowledgements}

Authors would like to thank Prof M. Pinar Menguc for his remarks made on the  manuscript which
improved the clarity of the paper. We also would like to thank the anonymous referee for his 
strong comments, suggestions and for his patience which improved the quality of the paper.


{}

\begin {appendix}
\section {}

Equation (29) can be solved numerically by using discrete space theory of radiative transfer
[60], [61], and [62]. We discritise the
equation (29) on the $[x_(i)- z(n)- \mu(m)]$ discrete mesh and this becomes
\begin{eqnarray}
&&{\bf M}_m \bigg[{\bf I}^+_{i, n+1}-{\bf I}^+_{i, n}\bigg] + \tau_{n+ \frac {1}{2}} \sigma_i^+
{\bf I}^+_{i, n+\frac {1}{2}} \nonumber \\
&=&\tau_{n+\frac {1}{2}} \bigg[ \sigma_a {{\bf B}^{\prime}_{\epsilon, i, {n+\frac {1}{2}} }} +
\sigma_{a}^{ff} {{\bf B}^{ff}_{em, i, {n+\frac {1}{2}} }} \bigg] {\bf h} \nonumber \\
&+&\frac {3}{64 \pi} \tau_{n+ \frac {1}{2}} \bigg[ {\bf R}^{++}_{i,i^\prime,{n+\frac {1}{2}}} \
{\bf a}^{++}_{i^\prime,{n+\frac {1}{2}}} {\bf CI}^{+}_{i^\prime,{n+\frac {1}{2}}} \nonumber \\
&+& {\bf R}^{+-}_{i,i^\prime,{n+\frac {1}{2}}} \
{\bf a}^{+-}_{i^\prime,{n+\frac {1}{2}}} {\bf CI}^{-}_{i^\prime,{n+\frac {1}{2}}} \bigg]
\end{eqnarray}

\begin{eqnarray}
&&{\bf M}_m \bigg[{\bf I}^-_{i, n}-{\bf I}^-_{i, n+1}\bigg] + \tau_{n+ \frac {1}{2}} \sigma_i^-
{\bf I}^-_{i, n+\frac {1}{2}} \nonumber \\
&=&\tau_{n+\frac {1}{2}} \bigg[ \sigma_{a,i} {{\bf B}^{\prime}_{\epsilon, i, {n+\frac {1}{2}} }} +
\sigma_{a,i}^{ff} {{\bf B}^{ff}_{em, i, {n+\frac {1}{2}} }} \bigg] {\bf h} \nonumber \\
&+&\frac {3}{64 \pi} \tau_{n+ \frac {1}{2}} \bigg[ {\bf R}^{-+}_{i,i^\prime,{n+\frac {1}{2}}} \
{\bf a}^{-+}_{i^\prime,{n+\frac {1}{2}}} {\bf CI}^{+}_{i^\prime,{n+\frac {1}{2}}} \nonumber \\
&+& {\bf R}^{--}_{i, i^\prime, {n+\frac {1}{2}}} \
{\bf a}^{--}_{i^\prime, {n+\frac {1}{2}}} {\bf CI}^{-}_{i^\prime,{n+\frac {1}{2}}} \bigg]
\end{eqnarray}

where
\begin{eqnarray}
{\bf B}^{\prime}_{em, i, {n+\frac {1}{2}}}=
\frac {{\bf B}_{e}(x_i,\beta)} {n_{e}, {n+\frac {1}{2}}\sigma_T} \nonumber
\end{eqnarray}
\begin{eqnarray}
{\bf M}_m=\bigg[\mu_m \delta_{mk}\bigg]
\end{eqnarray}
\begin{eqnarray}
{\bf C}=\bigg[c_m \delta_{mk}\bigg]
\end{eqnarray}
\begin{eqnarray}
{\bf h}=\bigg[1,1,1,\ldots \bigg]^{T}_{m}
\end{eqnarray}
T represents the transpose and $\mu$'s and C's are the roots and weights of angle quadrature and
\begin{eqnarray}
\tau_{n+\frac {1}{2}}= n_{\epsilon} \sigma_T \Delta Z
\end{eqnarray}
\begin{eqnarray}
{\bf I}^{\pm}_{i,n}={\bf I}(x_i, \pm \mu_m, Z_n)
\end{eqnarray}
\begin{eqnarray}
\sigma^{\pm}_i&=&\sigma(\pm x_i) \nonumber \\
&=&\sigma(x \pm \mu\beta)
\end{eqnarray}
\begin{eqnarray}
&&{\bf R}^{++}_{i,i^\prime, {n+\frac {1}{2}}}= \nonumber \\
&&\frac {1}{x_{i^2}} \bigg[\frac {x_i}{x_{i^\prime}}
+\frac {x_{i^\prime}} {x} -2 \bigg(\frac {1}{x_{i^\prime}}-\frac {1}{x_i}\bigg)+
\bigg( \frac {1} {x_{i^\prime}} -\frac {1}{x_i} \bigg)^2 \bigg]_{n+\frac {1}{2}}
\end{eqnarray}
Similarly ${\bf R}^{+-}, {\bf R}^{-+}$ and  ${\bf R}^{--}$. In case of static medium,
\begin{eqnarray}
{\bf R}^{++}={\bf R}^{+-}={\bf R}^{-+}={\bf R}^{--}
\end{eqnarray}
The subscript ${n+\frac {1}{2}}$ refers to the average of the layer with boundaries $Z_n$ and
$Z_{n+1}$. If we write
\begin{eqnarray}
{\bf M}&=&
\left(\begin{array} {cccc}
    {\bf M}_m &\quad &\quad &\quad  \\
    \quad &{\bf M}_m &\quad &\quad \\
    \quad  &\quad &\ddots &\quad \\
    \quad  &\quad &\quad  &\quad {\bf M}_m
\end{array}\right),\nonumber\\
{\bf C}&=&
\left(\begin{array} {cccc}
    {\bf C}^+_m &\quad &\quad &\quad  \\
    \quad &{\bf C}^+_m &\quad &\quad \\
    \quad  &\quad &\ddots &\quad \\
    \quad  &\quad &\quad  &\quad {\bf C}^+_m
\end{array}\right),
\end{eqnarray}
\begin{eqnarray}
{\bf W}_{kk^\prime}=a_ic_j=\frac {A_i R_{kk^\prime}c_j}{\sum R_{kk\prime}A_i c_j}
\quad\quad \quad \quad a_i=\frac {A_i R_{kk^\prime}} {\sum R_{kk^\prime} A_i c_j}
\end{eqnarray}
then equations (A.1) and (A.2) would become for ${\bf I}^\prime$ energy points.

\begin{eqnarray}
&&{\bf M}\bigg[{\bf I}^+_{n+1}-{\bf I}^+_{n}\bigg] + \tau_{n+ \frac {1}{2}} \sigma^+_{n+\frac {1}{2}}
{\bf I}^+_{n+\frac {1}{2}} \nonumber \\
&=&\tau_{n+\frac {1}{2}} {\bf S}^+_{n+\frac {1}{2}}
+\tau_{n+\frac {1}{2}} \eta \bigg[ {\bf R}^{++} {\bf W}^{++} {\bf I}^+ + {\bf R}^{+-} {\bf W}^{+-}
{\bf I}^- \bigg]_{n+\frac {1}{2}}
\end{eqnarray}
\begin{eqnarray}
&&{\bf M} \bigg[{\bf I}^-_{n}-{\bf I}^-_{n+1}\bigg] + \tau_{n+ \frac {1}{2}}
{\bf I}^-_{n+\frac {1}{2}} \nonumber \\
&=&\tau_{n+\frac {1}{2}} {\bf S}^-_{n+\frac {1}{2}}
+\tau_{n+\frac {1}{2}} \eta \bigg[ {\bf R}^{-+} {\bf W}^{-+} {\bf I}^+ + {\bf R}^{--} {\bf W}^{--}
{\bf I}^- \bigg]_{n+\frac {1}{2}}
\end{eqnarray}
where
\begin{eqnarray}
\eta= \frac {3} {64 \pi}
\end{eqnarray}
\begin{equation}
\bf I^\pm_n=
\left(\begin{array}{c}
I^{\pm}_{1,n} \\
I^{\pm}_{2,n} \\
I^{\pm}_{3,n} \\
\ldots \\
 I^{\pm}_{I^\prime,n} \end{array}\right)
\end{equation}
\begin{eqnarray}
{\bf S}_{n+\frac {1}{2}}=\bigg[\sigma_a {\bf B}^\prime_{em,i, {n+\frac {1}{2}}}+
\sigma^{ff}_{a}{\bf B}^{ff}_{emi, {n+\frac {1}{2}}}\bigg]
\left(\begin{array}{c}
1 \\
1 \\
1 \\
\ldots \\
1 \end{array}\right)_{I^\prime}
\end{eqnarray}
If we use "diamond scheme"  [62]
\begin{eqnarray}
{\bf I}^{\pm}_{n+\frac {1}{2}} = \frac {1}{2} \bigg({\bf I}^\pm_n+{\bf I}^\pm_{n+1}\bigg)
\end{eqnarray}
and we write the equation (A.13) and (A.14) in the form of interaction principle. We obtain

\begin{eqnarray}
&&\left(\begin{array}{cc}
X11    &   X12 \\
X21    &   X22 \\ \end{array}\right)
\left(\begin{array}{c}
I^+_{n+1} \\
\\
I^-_{n \ \ }  \end{array}\right) \nonumber\\
&&=\left(\begin{array}{cc}
Y11    &   Y12 \\
Y21    &   Y22 \\ \end{array}\right)
\left(\begin{array}{c}
I^+_{n} \\
\\
I^+_{n+1 \ \ }  \end{array}\right)\nonumber\\
&&+\tau
\left(\begin{array}{c}
S^+\\
S^-\\
\end{array}\right)
\end{eqnarray}

where
$$
X11=
{\bf M} +\displaystyle \frac {1}{2}\tau \bigg(\sigma^+-\eta R^{++} W^{++} \bigg)
$$
$$X12=
-\displaystyle \frac {1}{2}R^{+-}W^{+-}
$$
$$
X21=
 -\displaystyle \frac {1}{2} \tau R^{-+}W^{-+}
$$
$$
X22=
{\bf M} +\displaystyle \frac {1}{2}\tau \bigg(\sigma^-- R^{--} W^{--} \bigg)
$$
$$
Y11=
{\bf M} -\displaystyle \frac {1}{2}\tau \bigg(\sigma^+-{\frac {-1}{2}} R^{++} W^{++} \bigg)
$$
$$
Y12=
+\displaystyle \frac {1}{2} \tau R^{+-}W^{+-}
$$
$$
Y21=
 +\displaystyle \frac {1}{2} \tau R^{-+}W^{-+}
$$
$$
Y22=
{\bf M} -\displaystyle \frac {1}{2}\tau \bigg(\sigma^-- R^{--} W^{--} \bigg)
$$
We can derive the transmission and reflection matrices and the source vectors, from the
equation (A.19) by the
comparison with the interaction principle [62] and  are given in the Appendix.

To calculate the diffuse radiation ie.,  the specific intensities at each spacial point
in the medium,
${\bf I}^{\pm}_n$, we use the scheme described in  [63]and [62].
We estimate
${\bf I}^+_{n+1}, {\bf I}^-_{n}$ at each point which are given by,
\begin{eqnarray}
{\bf I}_{n+1}^+=r(1,n+1){\bf I}^-_{n+1}+{\bf V}^+_{n+\frac {1}{2}}
\end{eqnarray}
\begin{eqnarray}
{\bf I}_{n}^-=t(n,n+1){\bf I}^-_{n+1}+{\bf V}^-_{n+\frac {1}{2}}
\end{eqnarray}
where $r(1,n+1), t(n,n+1)$ are the diffuse reflection and transmission matrices at the internal points.
${\bf V}^+_{n+\frac {1}{2}}, {\bf V}^-_{n+\frac {1}{2}}$ are the source vectors which represent the
emission from the layer whose boundaries are $Z_n$ and $Z_{n+1}$ together with the diffuse radiation
from the rest of the medium.

The quantity ${\bf I}^+_{n+1}$ represents the reflected intensity in the direction of increasing
$\tau(\tau \rightarrow {\bf T}$, the total optical depth) while ${\bf I}^-_n$ represents the
emergent intensity
toward decreasing $\tau \rightarrow 0$. We can estimate these two quantities through the
equations (A.20) and
(A.21) at any spacial point in the medium in the form of angular distribution.

The transmission and reflection operators as are given by,
$$
{\mbox {\boldmath $t$}}(n+1, n) = {\bf G}^{+-} \bigl[\Delta^+\bf A^+ +g^{+-} + \bf g^{-+} \bigr]
$$
$$
{\mbox {\boldmath $r$}}(n+1, n) = {\bf G}^{-+} \bf g^{-+} \bigl[{\bf E}+\Delta ^+ {\bf A}^+ \bigr]
$$
$$
{\mbox {\boldmath $t$}}(n, n+1) = {\bf G}^{-+} \bigl[\Delta^-\bf A^- +\bf g^{-+} g^{+-} \bigr]
$$
$$
{\mbox {\boldmath $r$}}(n, n+1) = {\bf G}^{+-} \bf g^{+-} \bigl[{\bf E}+\Delta ^- {\bf A}^- \bigr]
$$
Where ${\bf E}$ is identity matrix
and
$$
{\bf G} ^{+-}=\bigl[{\bf E}-\bf g^{+-} \bf g^{-+}\bigr]^{-1}
$$
$$
{\bf G} ^{-+}=\bigl[{\bf E}-\bf g^{-+} \bf g^{+-}\bigr]^{-1}
$$
$$
\bf {g} ^{+-}=\frac {1}{2} \tau \Delta^+
$$
$$
\bf {g} ^{-+}=\frac {1}{2} \tau \Delta^- {\bf Y}
$$
$$
{\Delta} ^{+}=\bigl[{\bf M}+\frac {1}{2} \tau {\bf Z}\bigr]^{-1}; \quad {\bf Z}=\sigma -\eta {\bf R} {\bf W};
\quad {\bf Y}= \eta {\bf R}{\bf W}
$$

and the  source vectros are,
$$
\Sigma^+ = \tau {\bf G}^{+-} \bigl[\Delta^++\bf g^{+-} \bf \Delta^-\bigr] {\bf S}
$$
and
$$
\Sigma^- = \tau {\bf G}^{-+} \bigl[\Delta^-+\bf g^{-+} \bf \Delta^+\bigr] {\bf S}
$$
${\bf S}$ being the source function.
\end{appendix}
\end{document}